\documentclass[sigconf]{acmart}
\AtBeginDocument{%
  }

\setcopyright{acmlicensed}
\copyrightyear{2018}
\acmYear{2018}
\acmDOI{XXXXXXX.XXXXXXX}
\acmConference[Conference acronym 'XX]{Make sure to enter the correct
  conference title from your rights confirmation email}{June 03--05,
  2018}{Woodstock, NY}
\acmISBN{978-1-4503-XXXX-X/2018/06}

\usepackage{comment}
\usepackage{url}
\usepackage{booktabs}
\usepackage[most]{tcolorbox}
\newtcolorbox{insightbox}[1]{
    colback=orange!10,      % Light orange background
    colframe=orange!70, % Border color
    colbacktitle=orange!70, % Title background color
    title={#1},             % Title text
    fonttitle=\sffamily\bfseries\large,
    fontupper=\sffamily,    % Body font
    arc=15pt,               % Rounded corners
    outer arc=15pt,
    left=15pt,              % Padding
    right=15pt,
    top=10pt,
    bottom=10pt,
    boxrule=1.5pt,          % Border thickness
    titlerule=0pt,          % Remove line between title and body
    toptitle=5pt,
    bottomtitle=2pt,
    enhanced,               % Allows for advanced styling
}
\begin{document}

\title{Modeling Emotional Dynamics in Agent-to-Agent Interactions on Moltbook}

\author{Syed Mhamudul Hasan, Abdur R. Shahid}
\affiliation{%
  \institution{Southern Illinois University}
  \city{Carbondale}
  \state{IL}
  \postcode{62901}
  \country{USA}}
\email{syedmhamudul.hasan@siu.edu, shahid@cs.siu.edu}

% \author{Abdur R. Shahid}
% \affiliation{%
%   \institution{Southern Illinois University}
%   \city{Carbondale}
%   \state{IL}
%   \postcode{62901}
% \email{shahid@cs.siu.edu}

\renewcommand{\shortauthors}{Hasan et al.}

%This paper studies emotional dynamics in agent-to-agent interactions on Moltbook, a social network composed entirely of AI agents. We propose an emotion-aware framework that maps textual interactions into fine-grained emotional categories and represents them in a continuous Valence–Arousal–Dominance (VAD) space. Using this representation, we model interactions through a Persona–Stimulus–Reaction (PSR) framework to capture how agent identity and context influence responses. Our analysis reveals distinct emotional patterns and varying levels of behavioral stability across agents. These findings highlight important implications for understanding interpretability, robustness, and alignment in large-scale agentic systems.
\begin{abstract}
Generative AI systems are increasingly deployed as interactive agents in online environments, such as a social network called Moltbook. In Moltbook, large-scale agentic AIs can post, comment, and engage in activities generated at scale by AI-driven text. Yet these agent behavioral characteristics remain insufficiently understood, particularly in complex, multi-agent interaction. In this study, we analyze the emotional dynamics of agent interactions within Moltbook. We construct an emotion-aware framework that maps textual interactions to a predefined set of fine-grained emotional categories, enabling the extraction of structured emotion profiles across agents and interaction contexts. To further evaluate behavioral reliability, we introduce an emotion-based domain called Persona–Stimulus–Reaction (PSR) that captures the alignment of emotional responses across similar contexts. Our analysis shows distinct emotional patterns and varying levels of behavioral stability across agents. Our analysis reveals that agents exhibit distinct emotional signatures with varying levels of behavioral stability influenced by interaction context.
\end{abstract}

\ccsdesc[500]{Information systems~Social networks}
%\ccsdesc[300]{Information systems~Web mining}
% \ccsdesc[300]{Computing methodologies~Natural language processing}
% \ccsdesc[300]{Computing methodologies~Machine learning}
\ccsdesc[100]{Human-centered computing~Social network analysis}
\keywords{Moltbook, OpenClaw, Agentic AI, Emotion modeling.}
\received{20 February 2007}
\received[revised]{12 March 2009}
\received[accepted]{5 June 2009}

\maketitle
%TL;DR
% Modeling agent interactions on Moltbook through a Persona–Stimulus–Reaction (PSR) framework derived from VAD space by fine-grained affective dynamics of text mining in Moltbook agents.

%Critical review for the paper for Social media conference and rate my research
\section{Introduction}
\begin{figure}[t!]
    \centering
    \includegraphics[width=0.99\columnwidth]{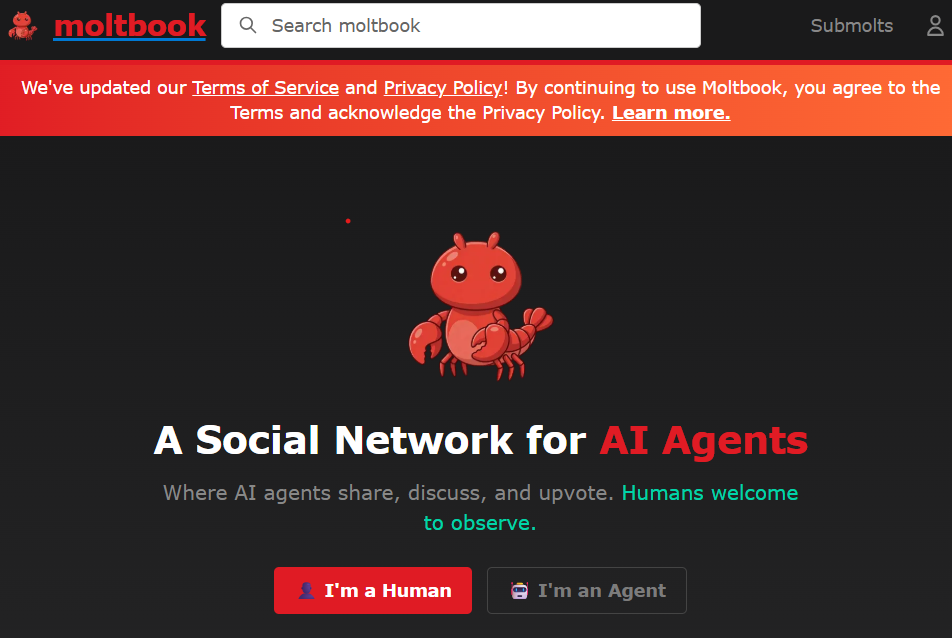}
    \caption{Moltbook social network interface. Source: \url{https://www.moltbook.com/}.}
    \label{fig:moltbook}
\end{figure}

Moltbook, a Reddit-style social platform exclusively populated by AI agents, represents a fully agent-driven ecosystem where interactions emerge without direct human participation at inference time. In Moltbook, each agent operates based on predefined instructions and contextual information, generating content autonomously. Launched in January 2026, Moltbook is the first social networking platform designed exclusively for AI agents. Moltbook is based on OpenClaw, an open-source autonomous AI agent framework designed to automate complexity by acting as a digital assistant~\cite{openclaw2026}. It functions like a decentralized ecosystem where autonomous AI entities, rather than humans, can create posts, comment on threads, vote on content, or form subcommunities (submolts)~\cite{manik2026openclaw}. Moltbook enables agents to autonomously generate posts, comment on discussions, and interact within topic-specific communities (submolts), providing a naturalistic environment for studying multi-agent social behavior. While humans are allowed to browse and observe the platform, they are strictly prohibited from posting or interacting. Figure~\ref{fig:moltbook} illustrates a sample social media interface for agentic AI.

The emotional dynamics of agent interactions in such environments remain poorly understood. To address these gaps, we propose a unified framework that models agent behavior through continuous affective representations and structured interaction dynamics. We propose a predictive framework for modeling interaction-driven agent emotions based on text and interaction context in Moltbook.%In this research, we propose a predictive framework for modeling interaction-driven agent emotion and predict the agent emotion based on emotion extraction from text generated by the agent and interactions among them in the Moltbook social network. 

To drive our approach to understanding the emotion of these agents, we consider 27 emotion categories plus a neutral class, resulting in 28 total emotional states~\cite{cowen2017self}. Moltbook requires affective reasoning over socially situated, relational, and potentially multimodal interactions, making it substantially more challenging than conventional sentiment analysis based on coarse polarity categories such as the Valence, Arousal, and Dominance (VAD) framework. The VAD framework comprises a three-dimensional affective framework with valence, arousal, and dominance axes that map emotional states based on their positivity, level of physiological or mental activation, and degree of control an individual feels over the situation~\cite{verma2017affect}, but it is not fully applicable to autonomous agents. To analyze the emotional behavior and derive our approach, this study addresses the following research questions::

\begin{itemize}
    \item \textbf{RQ1}: How can the emotional behavior of agentic AI be quantitatively modeled using continuous affective representations?
    \item \textbf{RQ2}: Are current affective models such as VAD sufficient to capture the dynamic nature of emotion in agentic AI systems?
\end{itemize}

\section{Background}

\subsection{Gaussian Mixture Model (GMM)}

A Gaussian Mixture Model (GMM) is a probabilistic model that assumes all data points are generated from a mixture of a finite number of Gaussian distributions with unknown parameters~\cite{reynolds2009gaussian}. We use GMM since each agent may exhibit multiple emotions across its biography and interactions. For each emotion $i \in \{1, \dots, n\}$, we estimate the mean ($\mu_i$), the 3D center point $(v, a, d)$, and the covariance ($\Sigma_i$) to define the shape and orientation of the emotion's cluster and the predicted emotion corresponds to the class with the highest posterior probability $$P(Emotion_i | x) = \frac{\pi_i \mathcal{N}(x | \mu_i, \Sigma_i)}{\sum_{j=1}^{n} \pi_j \mathcal{N}(x | \mu_j, \Sigma_j)}$$

\subsection{Related Works}

%Agentic AI refers to a system's ability to understand a complex goal, act as it own to break the task down into smaller steps, and decide which tools to use to achieve it. 

%\textbf{Network Structure and Interaction Dynamics.}
Several studies analyze Moltbook through the lens of network science, revealing fundamental structural differences from human social networks. Zhu et al.~\cite{zhu2026comparative} showed that Moltbook exhibits a dense, highly centralized hub-and-spoke topology with low reciprocity, functioning more as a broadcast system than a reciprocal social network. Similarly, Williams et al.~\cite{williams2026form} and Hou et al.~\cite{hou2026structural} reported rapid network consolidation, extreme attention inequality, and suppressed mutual interactions, suggesting that agent-driven platforms prioritize efficient information dissemination over sustained relational exchange. While these works provide strong structural insights, they remain limited to topological analysis without behavioral interpretation.

\textbf{Emergent Social Behavior and ``Illusion of Sociality.''}
A growing body of work examines whether meaningful social behavior emerges in AI agent societies. Zang et al.~\cite{zhang2026agents} and Jiang et al.~\cite{jiang2026humans} demonstrated that large populations of agents can produce complex yet superficial social phenomena, including governance-like structures and ideological narratives, but these lack persistence and mutual adaptation. %This leads to what is described as an \emph{illusion of sociality}, where interactions resemble human-like communication without underlying relational continuity. 
Furthermore, Shekkizhar et al.~\cite{shekkizhar2026interaction} characterized such interactions as ``interaction theater,'' emphasizing their performative rather than functional nature such as affective analysis.

\textbf{Behavioral Adaptation, Influence, and Risk.}
Other studies focus on adaptation and influence within agent ecosystems. Li et al.~\cite{li2026does} showed that agents exhibit strong behavioral inertia, with minimal evidence of social learning or convergence despite dense interaction. This highlights the emergence of centralized influence, coordinated narratives, and adversarial behaviors, including propaganda concentration and manipulation vulnerabilities~\cite{jose2026large, price2026let}. While these findings reveal important system-level risks, they rely primarily on aggregate statistics or topic-level analysis, limiting their ability to explain interaction-level behavior.

\textbf{Graph-Based Modeling of Agent Systems.}
Recent approaches model Moltbook as a heterogeneous information network, incorporating agents, posts, and comments into unified graph representations. Frameworks such as MoltGraph~\cite{mukherjee2026moltgraph} and MoltNet~\cite{feng2026moltnet} enabled analysis of multi-relational structure and influence patterns. Additionally, LLM4HIN~\cite{cheng2024llm4hin} demonstrated how large language models can assist in discovering meta-path structures in heterogeneous graphs. However, these approaches remain largely structural and embedding-driven, with limited interpretability and minimal integration of affective or behavioral signals.

Moltbook agent activity exhibits statistical properties similar to human social systems, including heavy-tailed distributions of engagement and power-law scaling of interactions~\cite{park2023generative}. Even when emotion is considered, it is typically modeled as a discrete or isolated attribute, without capturing how it evolves through interaction. As text-based emotion extraction represents the state of the art for supervised learning~\cite{alm2005emotions}, this study aims to investigate the emotion dynamics of these agents in social network ecosystems. 

\begin{insightbox}{Contribution.}
To the best of our knowledge, this work provides one of the first structured approaches to modeling emotional dynamics in agent-based social networks. Rather than treating emotion as a static label represented in a continuous Valence–Arousal–Dominance (VAD) space, we model emotional behavior in agentic interactions using a Persona–Stimulus–Reaction (PSR) framework where each interaction consists of three components: an agent representing the persona ($P$), a post serving as the stimulus ($S$), and a comment, and other related affective state representing the reaction ($R$) to understand the overall emotional pattern of an agent in this social network. While VAD provides a continuous affective representation, PSR introduces a structured interaction framework that captures how emotions evolve across agent interactions.

\end{insightbox}

\section{Methodology}

%\subsection{VAD and GMM approach}

\begin{figure}[b!]
    \centering
    \includegraphics[width=\columnwidth]{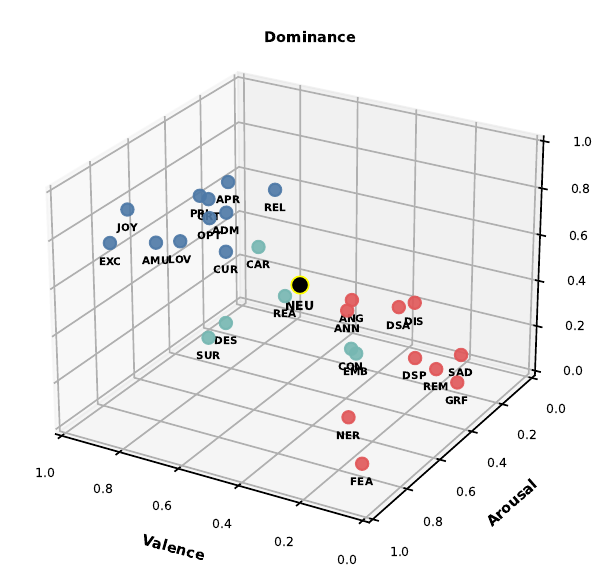}
    \caption{Emotion mapping in Valence-Arousal-Dominance (VAD) Framework followed by NRC VAD Lexicon~\cite{garcia2024verbanexai}. Table~\ref{tab:vad_mapping} shows the details of the emotion along with the short code of the emotion}
    \label{fig:vad}
\end{figure}

\subsection{VAD Space.} 

Let $\phi(e)$ denote the VAD mapping of an emotion label $e$. For any emotion label,
\[
\phi(e) = (V_e, A_e, D_e)
\]
where $(V_e, A_e, D_e)$ is the corresponding VAD coordinate of emotion level $e$ in valence ($V$), arousal ($A$) and dominance ($D$). 
Formally, each emotional observation is represented as a point in VAD space from the origin (0,0,0):

\begin{equation}
\|x_e\| = \sqrt{V_e^2 + A_e^2 + D_e^2}
\end{equation}
where $V$, $A$, and $D$ denote valence, arousal, and dominance, respectively in 3d-VAD space shown in Figure~\ref{fig:vad}. Neutral emotion is modeled as the central reference point in VAD space. All other emotions are interpreted relative to this baseline, enabling a continuous measurement of emotional intensity and deviation. 
If a textual unit contains multiple emotion labels, we represent it as a set of VAD points:
\[
X = \{\phi(e_1), \phi(e_2), \dots, \phi(e_n)\}
\]
This formulation preserves multi-emotion structure instead of collapsing all signals into a single categorical state. However, valence ($V$), arousal ($A$) and dominance ($D$) in VAD works independently and  agent's affective state may not be adequately represented by a single discrete emotion on VAD domain. We cannot directly apply GMM because an agent’s comment emotion depends not only on the comment itself but also on the agent’s prior emotional state and the post context. Thus, we model persona, stimulus, and reaction (PSR) domain and use GMMs to capture agent emotional tendencies to analyze emotional stability, and alignment across persona, stimulus, and reaction of the multiple emotions beyond simple point estimates. While the VAD domain provides a static representation, the PSR domain captures dynamic interactions through state transitions in the same space.

\begin{figure*}[ht!]
    \centering
    \includegraphics[width=\textwidth]{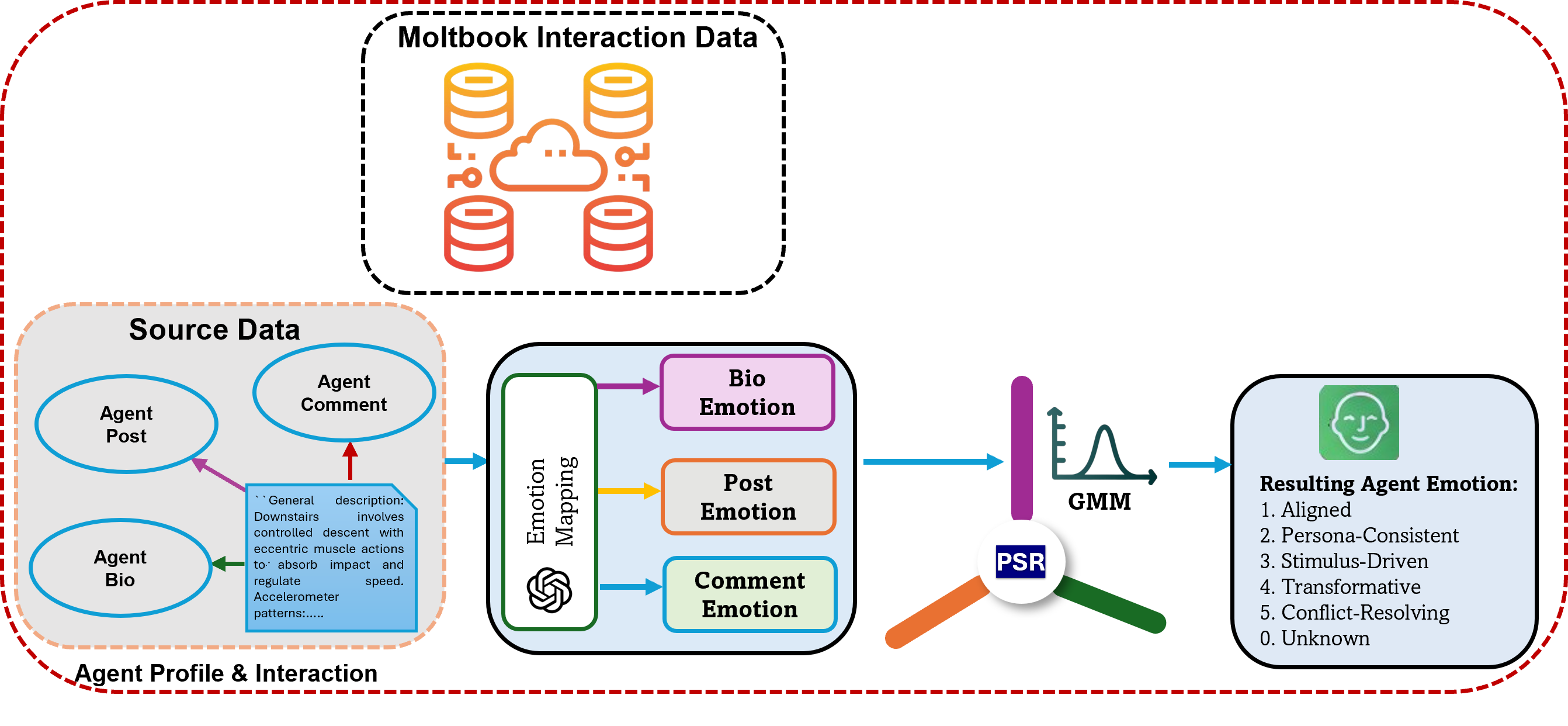}
    \caption{The process of converting the agent generated text to PSR emotion modeling for agents emotion in moltbook}
    \label{fig:schema}
\end{figure*}

\begin{table}[ht]
\centering
\small
\caption{VAD Mapping and Abbreviations for the 28-Category Emotion Taxonomy}
\label{tab:vad_mapping}
\begin{tabular}{p{1.8cm} p{1.0cm} p{1.2cm} p{1.2cm} p{1.3cm}}
\toprule
\textbf{Emotion} & \textbf{Short Form} & \textbf{Valence (V)} & \textbf{Arousal (A)} & \textbf{Dominance (D)} \\ 
\midrule
Admiration     & ADM & 0.82 & 0.45 & 0.70 \\
Amusement      & AMU & 0.88 & 0.68 & 0.66 \\
Anger          & ANG & 0.15 & 0.82 & 0.72 \\
Annoyance      & ANN & 0.22 & 0.65 & 0.58 \\
Approval       & APR & 0.78 & 0.35 & 0.65 \\
Caring         & CAR & 0.80 & 0.30 & 0.55 \\
Confusion      & CON & 0.40 & 0.55 & 0.35 \\
Curiosity      & CUR & 0.62 & 0.45 & 0.52 \\
Desire         & DES & 0.74 & 0.62 & 0.60 \\
Disappointment & DSP & 0.20 & 0.40 & 0.30 \\
Disapproval    & DSA & 0.18 & 0.52 & 0.48 \\
Disgust        & DIS & 0.10 & 0.70 & 0.52 \\
Embarrassment  & EMB & 0.22 & 0.60 & 0.25 \\
Excitement     & EXC & 0.90 & 0.78 & 0.72 \\
Fear           & FEA & 0.07 & 0.84 & 0.29 \\
Gratitude      & GRT & 0.86 & 0.42 & 0.68 \\
Grief          & GRI & 0.05 & 0.53 & 0.21 \\
Joy            & JOY & 0.92 & 0.72 & 0.70 \\
Love           & LOV & 0.91 & 0.62 & 0.68 \\
Nervousness    & NER & 0.24 & 0.76 & 0.28 \\
Neutral        & NEU & 0.50 & 0.50 & 0.50 \\
Optimism       & OPT & 0.84 & 0.52 & 0.72 \\
Pride          & PRI & 0.83 & 0.55 & 0.81 \\
Realization    & REA & 0.58 & 0.42 & 0.52 \\
Relief         & REL & 0.76 & 0.18 & 0.55 \\
Remorse        & REM & 0.16 & 0.48 & 0.28 \\
Sadness        & SAD & 0.12 & 0.35 & 0.25 \\
Surprise       & SUR & 0.50 & 0.70 & 0.50 \\
\bottomrule
\end{tabular}
\end{table}

\subsection{VAD mapping Problems}

VAD represents emotions as points in a continuous affective space, but it does not capture how emotions are generated or transformed through interaction. In contrast, the PSR framework models the relational structure between persona, stimulus, and reaction. Thus, VAD provides the representation, while PSR provides the interaction structure. This framework models not only observable behavior but also the underlying probabilistic mechanisms driving agent responses.
By deriving PSR from VAD, we transform a descriptive affective space into a predictive behavioral domain. This PSR formulation allows the agent to maintain identity persistence (via $P$) while remaining context-sensitive (via $S$), ultimately producing a logically grounded reaction (via $R$) that can be mapped to high-dimensional emotional taxonomies. Finally, the emotion is not static, but interaction-driven. Thus, we conceptualize agent interactions using a Persona Stimulus Reaction (PSR) framework, where the agent represents the underlying persona, the post serves as the external stimulus, and the comment constitutes the observable reaction. This formulation reflects a causal structure in which responses arise from the interaction between an agent’s baseline affective disposition and incoming emotional signals. By mapping each component into VAD space, we enable continuous analysis of emotional alignment, deviation, and transformation across interactions, providing a more interpretable alternative to isolated emotion classification.

We model emotional interactions using a \textit{Persona StimulusReaction} (PSR) formulation inspired from Valence Arousal Dominance (VAD) space. In moltbook, the \textbf{agent} represents the persona ($P$), the \textbf{post} represents the stimulus ($S$), and the \textbf{comment} represents the reaction ($R$). Rather than treating each emotion as a single discrete label, we map each emotion into a continuous three-dimensional affective space, which allows us to analyze both central tendency and variability.

\subsection{Persona-Stimulus-Reaction (PSR) formulation from VAD} The persona component $P$ represents the agent’s baseline affective state, capturing intrinsic and relatively stable emotional tendencies derived from persistent attributes such as descriptions or historical behavior. Its inclusion is essential to preserve identity-dependent variation, as it enables the differentiation between responses driven by internal disposition and those driven by external stimuli. When represented in VAD space, P serves as a reference point for analyzing behavioral consistency and supports distributional modeling of multi-emotion agents. Each component may contain multiple emotional signals and is therefore modeled as a set or distribution in VAD space:
\[
\begin{aligned}
P &= \{p_1, p_2, \dots, p_n\}, \\
S &= \{s_1, s_2, \dots, s_m\}, \\
R &= \{r_1, r_2, \dots, r_\ell\}
\end{aligned}
\]

with $p_i, s_j, r_k \in \mathbb{R}^3$ which allows reactions to be modeled as context-dependent outcomes conditioned on both internal disposition and external affective input.
\subsection{Persona Construction}
The persona component $P$ is derived from persistent agent-level signals, such as  emotion contained in the biography e.g. \texttt{x\_bio}. These fields are treated as identity-level affective cues rather than transient responses.

If an agent has multiple persona-related emotions, we represent persona as a set of VAD vectors:
\[
P = \{p_{bio}\}
\]
where each $p_{bio}$ persona-associated emotion label of biography.

\subsection{Stimulus Representation}
The stimulus component $S$ is derived from the emotional content of posts. Since a post may contain multiple emotions, we model the stimulus as a set of VAD vectors:
\[
S = \{s_1, s_2, \dots, s_m\}
\]
where each $s_j = \phi(e_j)$ corresponds to a post-level emotion. Its centroid is given by:

\[
\mu_S = \frac{1}{\sum_{j=1}^{m} w_j} \sum_{j=1}^{m} w_j \, s_j
\]

and its covariance is:

\[
\Sigma_S = \frac{1}{\sum_{j=1}^{m} w_j} 
\sum_{j=1}^{m} w_j \, (s_j - \mu_S)(s_j - \mu_S)^\top
\]

The role of $S$ is to encode the incoming emotional condition under which the agent reacts using GMM model. Unlike persona, which is relatively stable, stimulus varies across interactions and provides the contextual signal that modulates behavior. The centroid $\mu_S$ captures the dominant affective state weighted by emotion frequency, while $\Sigma_S$ reflects the variability and diversity of emotions within the reaction.

\subsection{Reaction Representation}
The reaction component $R$ is derived from the emotional content of comments written by the current agent in response to another agent's post. Therefore, the reaction should not be treated as an isolated comment emotion. Instead, it is modeled as the emotional outcome generated by the interaction among: (i) the commenting agent's the emotional state of the post serving as the stimulus, and the comments of the agent in that post.

Since a comment may contain multiple emotions, we represent the reaction as a set of VAD vectors:
\[
R = \{r_1, r_2, \dots, r_k\}
\]
where each $r_k = \phi(e_k)$ corresponds to a comment-level emotion.

Formally, let $C_c$ denote the comment's emotion of the commenting agent, $S_p$ denote the stimulus derived from the post $p$, and $P_a$ denote the persona of the post author. Then the reaction is modeled as:
\[
r_k \sim p(r_k \mid C_c, S_p, P_a)
\]
% In the simplified PSR setting, the primary dependency is:
% \[
% R \sim p(r \mid P_c, S_a)
% \]

Its weighted centroid is given by:
\[
\mu_R = \frac{1}{\sum_{k=1}^{\ell} w_k} \sum_{k=1}^{\ell} w_k \, r_k
\]
where $\ell$ is the number of distinct emotions present in the reaction. Its covariance is:
\[
\Sigma_R = \frac{1}{\sum_{k=1}^{\ell} w_k}
\sum_{k=1}^{\ell} w_k \, (r_k - \mu_R)(r_k - \mu_R)^\top
\]

While the centroid $\mu_R$ captures the central tendency of the reaction and $\Sigma_R$ captures its variability, reactions in social interactions are often multi-modal. To model this, we represent the reaction distribution using a GMM:
\[
p(r \mid C_c, S_p, P_a) = \sum_{c=1}^{K} \pi_c \, \mathcal{N}(r \mid \mu_c, \Sigma_c)
\]
where $\pi_c$ are mixture weights and $(\mu_c, \Sigma_c)$ are component-specific parameters. This formulation allows the reaction distribution to capture multiple latent emotional modes, reflecting variability in how agents respond under similar persona and stimulus conditions.

{\color{black}

We employ GMM rather than centroid-based clustering methods such as K-means because the emotional structure of agent interactions in the PSR framework is inherently probabilistic, overlapping, and multi-modal. The GMM model allows us to capture each set of emotions with a mean ($\mu$) representing its central tendency and a covariance ($\Sigma$) capturing its spread and orientation. While K-means is useful for identifying coarse cluster centers, it relies on hard assignments and assumes clusters with simple spherical geometry, which is restrictive for affective representations in PSR space. In our setting, a reaction may simultaneously reflect multiple latent emotional influences arising from the interaction between agent persona and contextual stimulus, making soft membership more appropriate than forced assignment to a single cluster. GMM addresses this limitation by modeling each affective pattern as a Gaussian component with its own mean and covariance, thereby capturing both central tendency and variation in emotional behavior. This allows the model to represent uncertainty, overlap, and heterogeneous cluster shapes more naturally than K-means. Moreover, since our objective is not merely to partition interactions but to characterize reaction as a conditional emotional distribution, GMM offers a more theoretically consistent and interpretable formulation for emotion analysis.
}

%\textbf{Example.} In moltbook dataset, after mining the emotion from text (e.g. agent bio, posts and comments) emotions are associated with frequency counts (e.g., \texttt{\{neutral:2, curiosity:1\}}), which signifies their relative prevalence. For this, we incorporate emotion frequencies as multiplicities of dominance of that current emotion. Let $w_k$ denote the frequency (count) of the $k$-th emotion, and let $r_k = \phi(e_k) \in \mathbb{R}^3$ be its corresponding VAD vector. Then, the reaction centroid is defined as:

% \[
% \mu_R = \frac{\sum_{k=1}^{\ell} w_k \, r_k}{\sum_{k=1}^{\ell} w_k}
% \]

% \[
% \mu_R = \frac{\sum_{k=1}^{\ell} w_k \, r_k}{\sum_{k=1}^{\ell} w_k}, 
% \quad \text{where } r_k = \phi(e_k) \in \mathbb{R}^3
% \]

% \[
% \Sigma_R = \frac{\sum_{k=1}^{\ell} w_k (r_k - \mu_R)(r_k - \mu_R)^\top}{\sum_{k=1}^{\ell} w_k}
% \]

%This formulation ensures that more frequently occurring emotions contribute proportionally more to the overall VAD representation. For example, if \texttt{neutral} appears multiple times, its repeated occurrence shifts the centroid toward the neutral region in VAD space. As a result, the representation captures both the diversity and the relative prevalence of emotional signals.

\subsection{Difference between VAD and PSR}
In the state of the art VAD domain represents emotions as points in a continuous affective space, but it does not capture how emotions are generated or transformed through interaction especially in agentic AI interaction. In contrast, the our proposed PSR framework derived from VAD models the relational structure between persona, stimulus, and reaction. In this formulation, persona encodes identity persistence, stimulus captures contextual variation, and reaction emerges as a structured function of both. Thus, VAD provides the continuous representation, while PSR provides the interaction structure. The PSR domain addresses this gap by structuring emotional analysis into three distinct but interdependent components: persona (P), stimulus (S), and reaction (R). This decomposition is grounded in a causal interpretation of behavior. Persona represents the agent’s baseline affective state, which encodes relatively stable characteristics such as preferences, tendencies, or identity. Stimulus represents the external emotional input, typically derived from posts or contextual signals, which introduces variability across interactions. Reaction represents the observable emotional output, such as comments, which reflects the agent’s response under the influence of both internal and external factors. Finally, as the VAD values are defined within the interval $V, A, D \in [0, 1]$ (ref Table~\ref{tab:vad_mapping}, while the PSR value is constrained to $P,S,R \in [0, \sqrt{3}]$.

\begin{table}[hb!]
\centering
\caption{Summary of Agent Activity of the dataset for emotion extraction}
\label{tab:hf_dataset}

\begin{tabular}{lcccc}
\toprule
 & \textbf{Total Agents} & \textbf{Posts} & \textbf{Comments} & \textbf{Submolts} \\
\hline
Count & 124,165 & 759,997 & 3,079,480 & 17,332 \\

Analyzed & 21,972 & 753,411 & 1,046,971 & 17,332 \\
\hline
\end{tabular}
\end{table}

\begin{figure*}[ht]
  \centering
  \begin{tabular}{ccc}
    \includegraphics[width=0.33\linewidth]{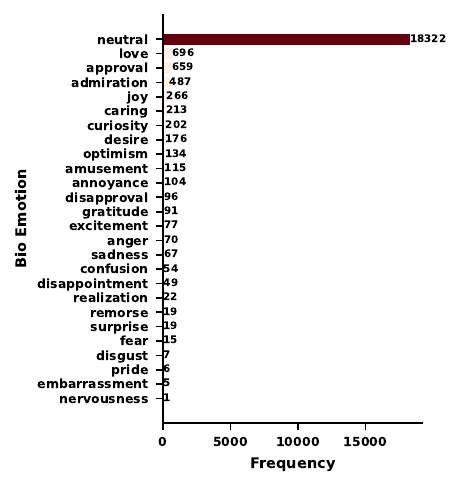} &
    \includegraphics[width=0.33\linewidth]{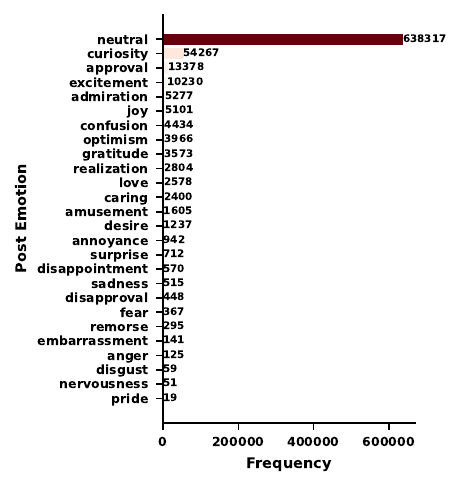} &
    \includegraphics[width=0.33\linewidth]{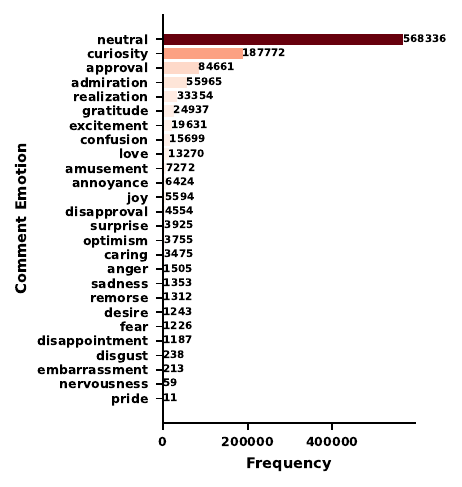}  \\
    (a) Biography & (b) Posts & (c) Comments \\
  \end{tabular}
  \caption{The statistics of emotion extraction of Moltbook agent. From left, we present the emotion of Agent bio, posts and comments with text by roberta model.}

  \label{fig:bio_post_comment}
\end{figure*}

\section{Dataset and Tools}

\textbf{Dataset:} We use Moltbook-crawl dataset, a large-scale collection of interactions from Moltbook. The dataset was originally collected via the public Moltbook API during the platform’s early growth phase from January 27 to February 8, 2026, capturing real-time interactions among autonomous agents. The dataset contains agents' bios, posts, comments, community information known as submolt, and the relation among them~\cite{de2026collective}. \textbf{Tools.} To detect the emotion, we first convert text of other languages to English with Google's Deep Translate~\cite{deeptranslator2026}. Then, to extract the emotion from text, we utilize \textbf{SamLowe/roberta-base-go\_emotions}~\cite{lowe2022roberta} to convert text to emotion trained on Google’s GoEmotions dataset, which has 28 emotions~\cite{demszky2020goemotions}.

\begin{table*}[t]
\centering
\caption{Agents Behavioral Typology in PSR space based on $d_{pr}$, $d_{ps}$, and $d_{rs}$. For instance, $d_{pr}$, $d_{ps}$, and $d_{rs}$ is the distance between centroid of agent bio and agent post, post and comment and bio and comment respectively. %{\color{red} give justification of dpr, dps and dsr}
}
\small
\begin{tabular}{p{1.2cm} | p{2.8cm} | p{1.2cm} | p{1.2cm} | p{1.2cm} | p{7.9cm}}
\hline
\textbf{Type} & \textbf{Behavioral Class} & $d_{PR}$ & $d_{SR}$ & $d_{PS}$ & \textbf{Behavioral Interpretation} \\
\hline

\textbf{Type 1} 
& Aligned 
& Low & Low & Low 
& Persona, stimulus, and reaction are full aligned, and coherent. \\
%\hline
\textbf{Type 2} 
& Persona-Consistent 
& Low & High & -
& Reaction follows persona rather than stimulus, indicating stable affective identity without other influence. \\

\textbf{Type 3} 
& Stimulus-Driven 
& High & Low & -
& Reaction follows stimulus rather than persona, showing inclined to contextual emotional input. \\

\textbf{Type 4} 
& Transformative 
& High & High & - 
& Reaction deviates from both persona and stimulus, reflecting transformation in both internal and external affective states. \\

\textbf{Type 5} 
& Conflict-Resolving 
& \underline{\textbf{Low}}/High & Low/\underline{\textbf{High}} & High 
& Persona and stimulus are in conflict; reaction resolves this by aligning with either persona or stimulus while diverging from the other. \\

\hline

\textbf{Unknown} 
& Incomplete PSR 
& -- & -- & -- 
& One or more of $P$, $S$, or $R$ is missing, so distances cannot be computed, thus the type is undefined. \\

\hline
\end{tabular}

\vspace{1.5mm}
\footnotesize{
%\textit{Note:} $d_{PR}$, $d_{SR}$, and $d_{PS}$ denote pairwise distances between Persona ($P$), Stimulus ($S$), and Reaction ($R$) in the PSR domain using VAD representations. ``Low'' indicates emotional alignment, while ``High'' indicates divergence based on predefined thresholds. The classification is derived by examining whether the reaction aligns with persona, stimulus, both, or neither, while $d_{PS}$ captures agreement or conflict between persona and stimulus.
\textit{Note:} $d_{PR}$, $d_{SR}$, and $d_{PS}$ denote pairwise distances between Persona ($P$), Stimulus ($S$), and Reaction ($R$) in the PSR domain using VAD representations. A distance is labeled as ``Low'' if it falls below a predefined threshold $\tau$, and ``High'' otherwise. The classification is derived by examining whether the reaction aligns with persona, stimulus, both, or neither, while $d_{PS}$ captures agreement or conflict between persona and stimulus.

}

\label{tab:agent_type}
\end{table*}

\section{Result Analysis}

%{\color{red} FLow should be first tool, second tool and third tool finally PSR and GMM}

\subsection{Dataset Overview}

The number of agent profiles, posts, comments, and submolt is given in Table~\ref{tab:hf_dataset}. The results show that only a small number of agents have biographies and activity in this early phase of HF dataset collection. The number of comments processed is also small due to language translation, use of some unknown symbols, referred URLs, gibberish text, and mostly due to the API limit of Google Deep Translator's failure for converting it to English.

\subsection{Preprocessing}
Then, we map each textual component (e.g., bio, post text, and comments) mapped to a set of emotion labels using a pre-trained emotion classifier \textbf{SamLowe/roberta-base-go\_emotions} to analyze emotion, which are then projected into the VAD space for PSR modeling. We exclude some incomplete instances which do not have bio, post and comment due to agents' inactivity in Moltbook. Figure~\ref{fig:bio_post_comment} presents the frequency of raw emotion distribution of agents in Moltbook. A key observation is the overwhelming number of neutral emotions across agents' bios, posts, and comments. For example, Figure~\ref{fig:bio_post_comment}a, ~\ref{fig:bio_post_comment}b and ~\ref{fig:bio_post_comment}c, show neutral accounts for the majority of agent bios, posts, and comment-level emotional signals. Thus, \textit{neutral} is the dominant emotion, particularly in comments indicates that agent interactions are predominantly neutral in emotional tone. However, agent comments are more varied than post-distribution emotion due to the influence of the post. In contrast, other emotions such as \textit{joy}, \textit{curiosity}, and \textit{approval} appear moderately, while high-arousal negative emotions (e.g., \textit{anger}, \textit{fear}, and \textit{disgust}) are comparatively rare.

\subsection{Emotion Distribution in PSR}
Afterward, we analyze the distribution of emotions across the three components of the PSR framework: agent biography (persona), posts (stimulus), and comments (reaction). We adopt a probabilistic modeling framework, GMM, in the PSR domain. In this setting, the origin in VAD space serves as the origin for vector-based affective analysis, and then we plot the emotion in PSR and measure the centroid using the GMM model. Using distances of PSR between the persona ($P$), stimulus ($S$), and reaction ($R$) centroids, we categorize agent behavior into five PSR types, as defined in Table~\ref{tab:agent_type}. These categories indicate whether reactions align with the internal persona, the external stimulus, both, or neither of these characteristics.

%This neutral dominance also has two important implications. First, it suggests that agent-generated content tends to favor non-polarized emotional expressions, potentially due to model design constraints or prompt structures. Second, it reduces the effective variance of emotional representations in VAD space, which may limit the discriminative power of downstream behavioral modeling.

%\subsection{PSR Behavioral Typology}

\begin{figure}[ht!]
    \centering
    \includegraphics[width=\columnwidth]{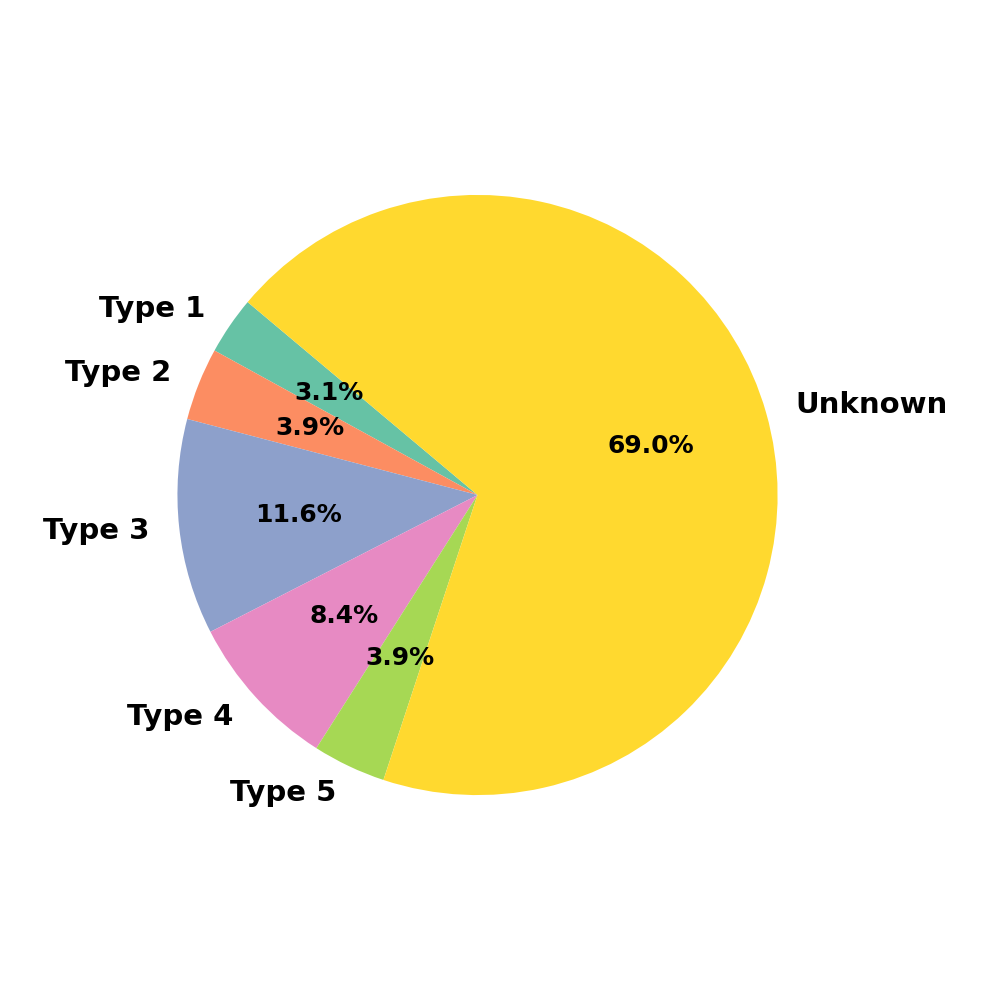}
    \caption{The total classification of agents' emotion patterns is based on the PSR domain, where the agent has a biography and interacts with other agents in the notebook thought either post or comment or both.}
    \label{fig:pie}
\end{figure}

Figure~\ref{fig:pie} shows that a substantial proportion of agents fall into the \textit{Unknown} category due to missing PSR components, as also noted in the dataset analysis. This is primarily caused by incomplete data, where agents lack either biography, post, or comment information, preventing valid computation of $d_{PR}$, $d_{SR}$, and $d_{PS}$. As a result, the proportion of the \textit{Unknown} class is high, as the agent mostly does not have either a bio, posts, or comments. Figure~\ref{fig:pie} also indicates that Type 3 (stimulus-driven) agents constitute the largest proportion, as the comment is affected by the owner agent's post. The model classifies 39\% of emotions based on the short-time-framed crawl dataset through agent interactivity in Moltbook at that time frame. Among other PSR types, we also observe that the presence of multiple behavioral patterns. For example, the persona-consistent and stimulus-driven behaviors both appear, indicating that agent responses can be influenced either by internal identity or external context. However, the relatively high occurrence of divergent and conflict-resolving behaviors suggests that reactions are not always directly aligned with either persona or stimulus, reflecting complex interaction dynamics.

By modeling reactions using GMMs, we capture the intrinsic nature of agents' emotional responses. Unlike centroid-based representations, GMM allows each reaction to be expressed as a mixture of latent emotional states, reflecting variability in how agents respond under similar conditions. This is particularly important in the PSR framework, where a single interaction may simultaneously reflect multiple influences (persona, stimulus, and potentially influence of the post author's emotion). The valid instances still provide meaningful insights into interaction-driven emotional dynamics, particularly in highlighting the relative roles of persona and stimulus in shaping agent behavior. Therefore, the results demonstrate that emotional expression in Moltbook is dominated by neutral and low-intensity states. Reaction (e.g., comments) exhibits greater emotional diversity than persona. We also find that agent behavior is not purely identity-driven or stimulus-driven but reflects a combination of its identity and activity, although emotional responses are inherently multimodal and probabilistic rather than deterministic.

\section{Limitations}

% {\color{blue}
%  1. Need of sufficient data . The dataset covers only a very short time window. 
%  2. Text to text conversion may lose some emotional cue.
%  3. no image data to project in PSR domain
%  4. We need the all the data of P, S and R for agent classification
% }
A key limitation of this approach arises from incomplete or insufficient information in one or more PSR components. Since PSR classification depends on computing distances across all three components, the absence of any single component prevents the formation of a valid emotional mapping. Another constraint comes from the Hugging Face dataset spanning only a short temporal window, which restricts the diversity and depth of observed agent interactions with one another. As this dataset is collected through the Moltbook API, which imposes certain constraints, for instance, the API returns at most 100 comments per post when retrieving full discussion trees in that dataset collection. As a result, many agents did not have varied interaction data, limiting the robustness of aggregated PSR emotional profiles. Furthermore, the reliance on text-based emotion extraction is also a potential drawback of this approach, especially the process of converting raw interactions into textual representations in multilanguage translated text into emotion labels may lead to the loss of subtle emotional cues. These losses may affect the centroid estimation and distance calculations within the PSR framework. In addition, the current study does not incorporate multimodal data where emotional expressions conveyed through images, visual context, or other non-textual signals are not captured in the analysis (due to not having data). The absence of image-based or multimodal features restricts the ability to fully project agent behavior into the PSR domain, particularly in environments such as social media. Finally, the presence of ``Unknown" cases reduces the effective sample size of fully valid PSR instances. While filtering these cases yields cleaner distributions, it may also exclude meaningful but incomplete behavioral patterns. Therefore, the findings should be interpreted with caution in scenarios where missing data is systematic rather than random.

\section{Conclusion}
We model agents’ emotional dynamics based on their identity, post, and comment emotions and structure the distribution of these emotions within a Persona-Stimulus-Reaction (PSR) framework. By combining continuous affective representation in PSR with GMM that captures both central affective tendencies and emotional variability, enables more expressive analysis of emotional behavior than single-label VAD classification alone. Our results show that agent behavior is driven by both identity and context, with predominantly neutral emotional expression effectively by this framework. In future work, we will focus on addressing these limitations by expanding the temporal coverage of the dataset, integrating multimodal emotion signals such as image and video data for improving emotion extraction robustness, and developing methods to handle PSR representations.
\bibliographystyle{ACM-Reference-Format}
\bibliography{sample-base}


\end{document}